\documentclass[letterpaper,floatfix,superscriptaddress,preprint]{revtex4}
\usepackage{latexsym}
\usepackage{amstext,amsfonts}
\usepackage{epsfig}

\begin{document}

\title{Dynamic networks and directed percolation (28 January 2009)}

\author{Roni Parshani}
\affiliation{Minerva Center \& Department of Physics, Bar-Ilan University,
Ramat Gan, Israel}
\author{Mark Dickison}
\affiliation{Center for Polymer Studies, Boston University, Boston,
Massachusetts 02215, USA}
\author{Reuven Cohen}
\affiliation{Department of Mathematics, Bar-Ilan University, Ramat Gan,
Israel}
\author{H. Eugene Stanley}
\affiliation{Center for Polymer Studies, Boston University, Boston,
Massachusetts 02215, USA}
\author{Shlomo Havlin}
\affiliation{Minerva Center \& Department of Physics, Bar-Ilan University,
Ramat Gan, Israel}

\date{28 January 2009 --- pdcsh.tex}

\begin{abstract}

We introduce a model for dynamic networks, where the links or the strengths of
the links change over time. We solve the model by mapping dynamic
networks to the problem of directed percolation, where the direction
corresponds to the evolution of the network in time. We show that the
dynamic network undergoes a percolation phase transition at a critical
concentration $p_c$, which decreases with the rate $r$ at which the
network links are changed. The behavior near criticality is universal
and independent of $r$. We find fundamental network laws are changed. (i) For Erd\H{o}s-R\'{e}nyi networks we find that the size of the giant component at criticality scales with the network
size $N$ for all values of $r$, rather than as $N^{2/3}$. (ii) In the presence of a broad distribution of disorder, the optimal path length between two nodes in a dynamic network scales as
$N^{1/2}$, compared to $N^{1/3}$ in a static network.

\end{abstract}
\maketitle

Network theory has answered many questions concerning static networks
\cite{Albert,Barabasi,Cohen,Pastor,Newman}
but many real networks are dynamic in the sense that their links, or
the strengths of their links, change with time. For example, in social
networks friendships are formed and dissolved, while in communication
networks, such as the Internet, the load (weight) on the links changes
continually. The challenges posed by such dynamic networks are
beginning to be addressed. For example, Kempe {\it et al.} have studied
algorithms for broadcasting or gossiping in dynamic networks
\cite{Kempe} while Volz and Meyers have studied the epidemic SIR
model on dynamic networks \cite{Volz}.

Fundamental questions that have been extensively studied in static
networks are still open for dynamic networks.  Here we ask: (i) Is there
a critical concentration of links for which the dynamic network
undergoes a percolation phase transition, above which order $N$ of the
network nodes are still connected and below which the network breaks
into small clusters? (ii) If so, what is the percolation threshold for
which the transition occurs, and how does it depend on the dynamics?
(iii) What are the properties near criticality?

\begin{figure}[h]
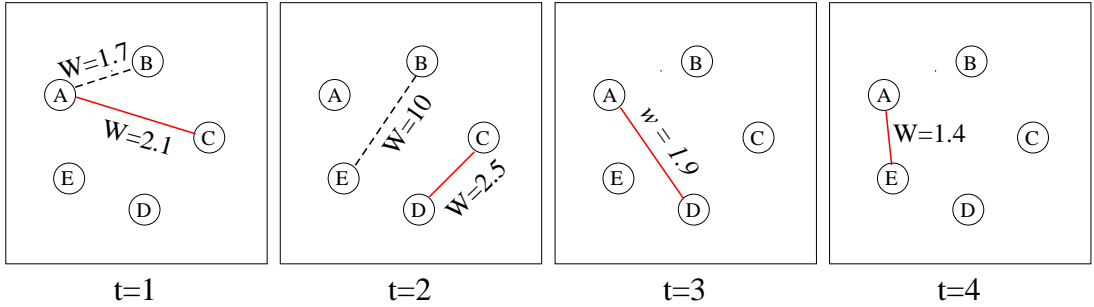

\begin{center}
\epsfig{file=Fig6.eps,height=4cm}
\epsfig{file=Fig7.eps,height=4cm}
\epsfig{file=Fig8.eps,height=4cm}
\epsfig{file=Fig9.eps,height=4cm}
\end{center}
\caption{A five-node dynamic network at different time steps. The weights
$W$ are defined in the text.}
\label{DNSteps}
\end{figure}

Consider an $N$-node network with $M$ links that change over
time. Each link has a lifetime $\tau$ drawn from a Poisson distribution,
and is replaced by a new link between two randomly-selected nodes after
its lifetime $\tau$ expires (Fig.~\ref{DNSteps}).  We define a unit time
step as the time required for a walker to traverse one link; we also
assume that when traversing the network one can only remain for a
limited time at each node. Even if there is no path between nodes $A$
and $B$ at a specific time, a walker traversing the network may be able
to pass from point A to point B because new links are continually
appearing. Likewise, even if a path between A and B exists at a given
time it may be disconnected before a walker is able to traverse it
(Fig.~\ref{DNSteps}).

\begin{figure}[h]
\begin{center}
\epsfig{file=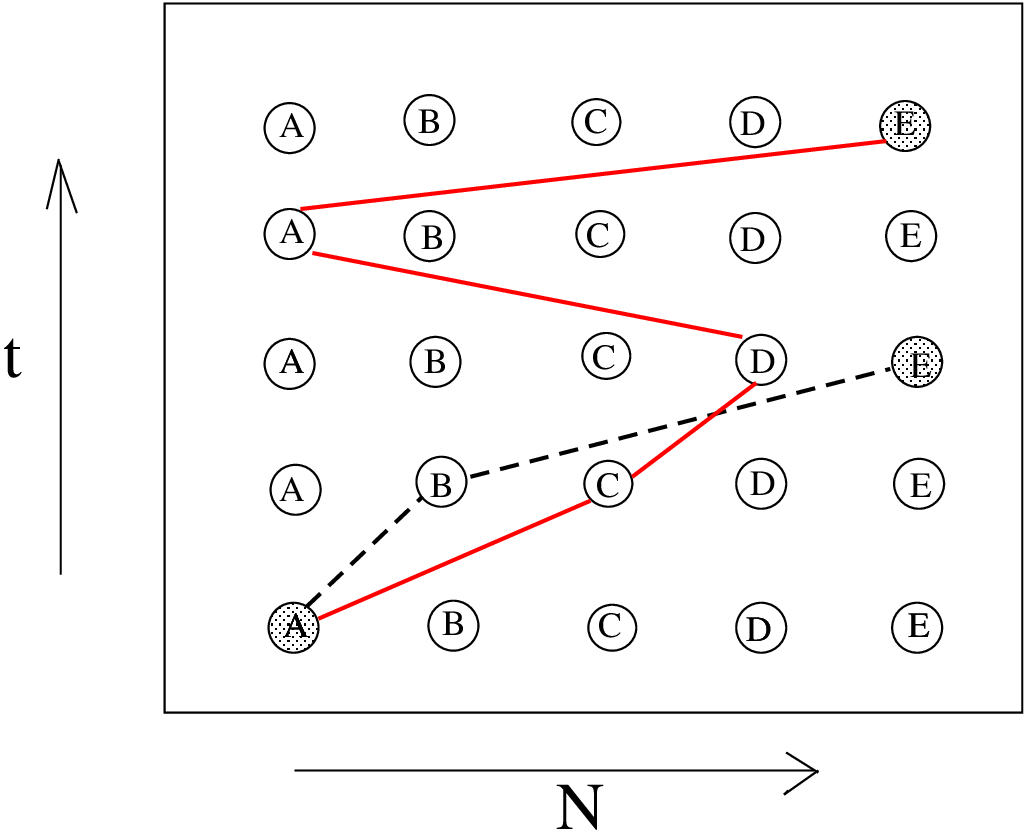,width=8cm,height=8cm}
\epsfig{file=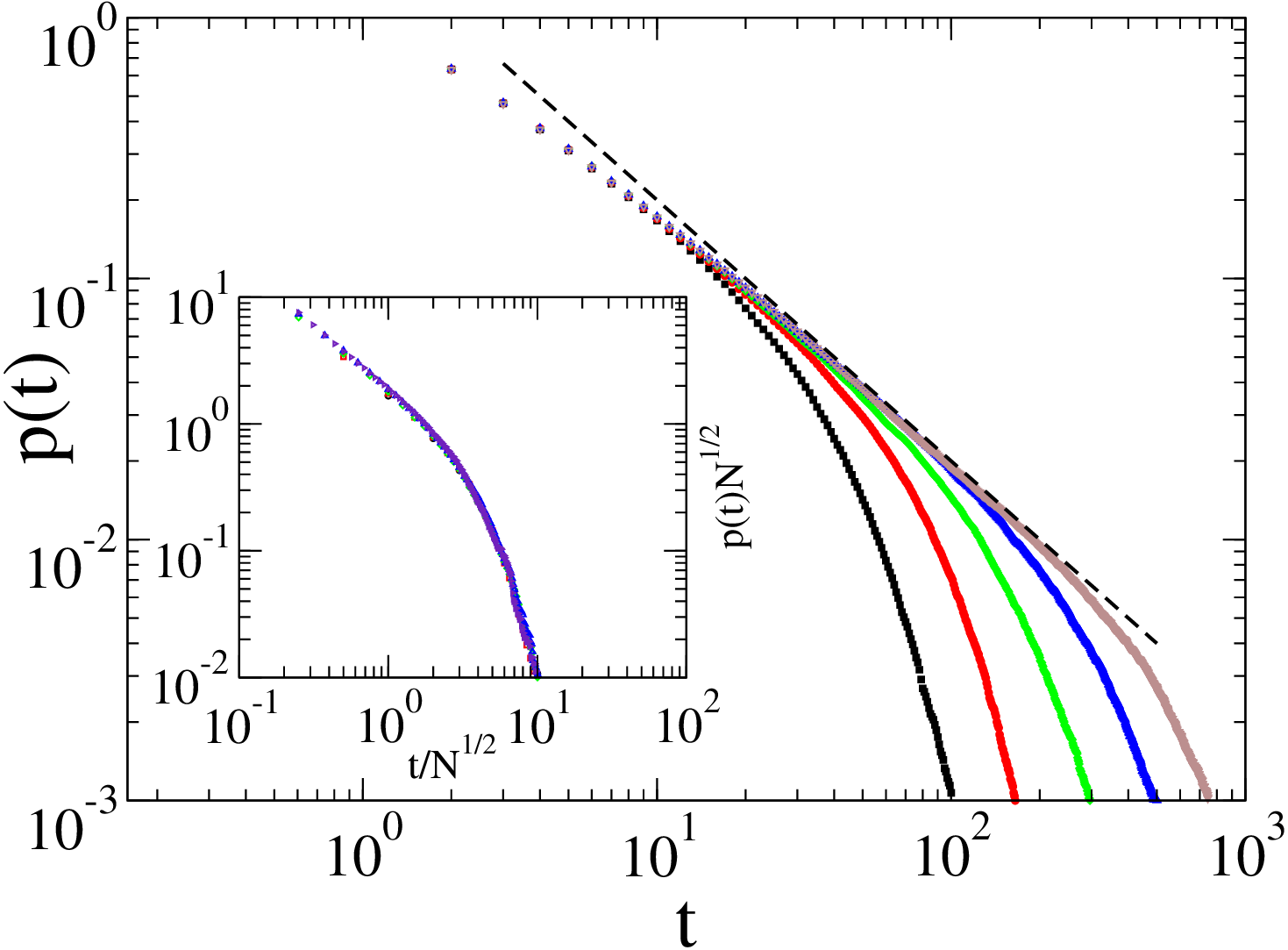,width=8cm,height=8cm}
\end{center}
\caption{(Color online) (a) The y axis corresponds to the time dimension
  and each successive layer of two rows corresponds to a network
  configuration at a different time step (c.f. Fig.1.) The full red line path is the optimal path (sum of its weights is
  minimal) between node A and node E, even though a shorter path exists,
  shown as a black dashed line.  (b) Simulation results for the
  survivability $P_s(t)$ and its cutoff, at criticality, for dynamic
  networks of different network sizes. From left to right: $N=100$,
  $400$, $1600$, $6400$, $25600$. The data collapse shown in the inset demonstrates that
 $P_s(t)$ in dynamic networks is universal when scaled by $N^{1/2}$.  }\label{DNFullAndSurvivelity}
\end{figure}

In order to represent the time evolution of the network we add to our
network representation another axis, which corresponds to a time axis.
Each layer along the time dimension corresponds to a network state at
some time $t$ [Fig.~\ref{DNFullAndSurvivelity}(a)]. Our new network
representation includes all the information regarding the network
dynamics.

We now argue that percolation on a dynamic network is equivalent to the
problem of directed percolation (DP) in infinite dimensions
\cite{DP,DP2}.  A percolation process can be described as an invasion of
a liquid in a porous media where the flow between two neighboring pores
is permitted with some probability $p$ and blocked with probability
$(1-p)$. This process undergoes a phase transition at the critical
probability $p=p_c$. For $p>p_c$ the liquid will spread throughout the
media, while for $p<p_c$ the liquid will be constrained to a finite
area. At exactly $p_c$ the liquid fills the incipient infinite cluster
(called the giant component in network theory). Directed percolation (DP), a special
kind of percolation, is a process in which the spread is limited to a
direction that is defined to be the longitudinal axis \cite{DP,DP2}.
While non-directed percolation has been successfully applied to networks
\cite{Cohen,Callaway,Percolation_SF,Percolation_directed_SF} and is
widely used for studying network stability
\cite{Cohen,Callaway,LPP,Cohen-3}, the analog of DP in networks has not been previously studied.

In a dynamic network the direction has the following interpretation. The evolution of the
network in time plays the role of the direction and the additional time
axis presented in our model acts as an additional longitudinal
(vertical) axis that is the hallmark of DP
[Fig.~\ref{DNFullAndSurvivelity}(a) and Fig.~\ref{DNSteps}]. The correspondence between percolation on
dynamic networks and DP on regular networks allows us to apply the
results known from DP at criticality to dynamic networks.

\begin{figure}[h]
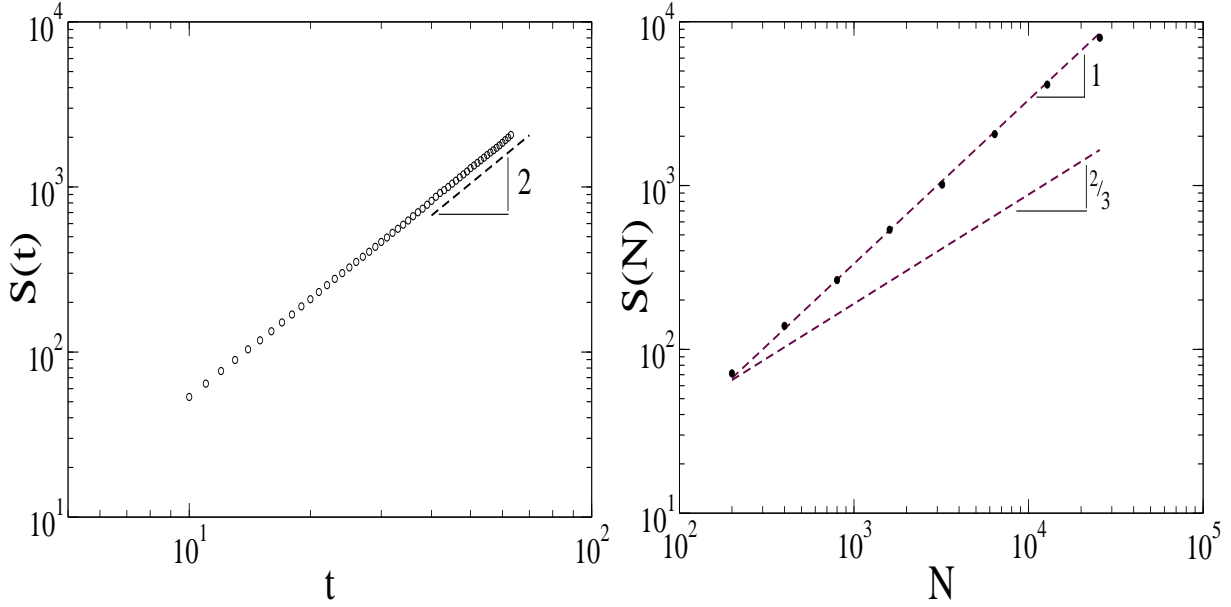

\begin{center}
\epsfig{file=Fig3.eps,height=8cm,width=8cm}
\epsfig{file=Fig4.eps,height=8cm,width=8cm}
\end{center}
\caption{(a) The cluster size $S(t)$ scales as $t^2$.  (b) Simulation
  results (dots) supporting the relation $S(N) \sim N$ (Eq.~3) (upper
  dashed line), the cluster size in dynamic networks, compared to the
  known $S(N) \sim N^{2/3}$ (bottom dashed line), the cluster size in
  static networks.} \label{ClusterSizeFig}
\end{figure}

Networks can be regarded as infinite dimensional structures since no
spatial constraints exist. Therefore we expect the critical properties
of dynamic networks to be the same as DP in infinite dimensions. The
relevant critical properties for DP are \cite{DP,DP2}: (i) $D(t)$, the
number of nodes reached at time $t$, scales as $D(t) \sim t$; $S(t)$ (ii) The giant component
size, scales as $S(t) \sim t^2$; and (iii) $P_s(t)$, the survivability (the
probability of reaching layer $t$ when growing a cluster), scales as
$P_s(t) \sim t^{-1}$.  Figures~\ref{DNFullAndSurvivelity}(b) and
\ref{ClusterSizeFig}(a) present simulation results confirming these scaling relations.

To learn about the size-dependent properties of dynamic networks we
determine the DP properties as a function of the network size $N$,
rather than as a function of $t$. In DP at criticality, the infinite dimensional relationship between $w$, the width in the transverse axes, and $t$, the length in the longitudinal axes, is  $w \sim t^{1/2}$.  The upper
critical dimension $d_c$ is the lowest dimension for which the system has the properties of an infinite dimensional system. For DP this value is $d_c =
4 + 1$ ($1$ corresponds to the longitudinal axis), so the relation between the system size at the upper critical
dimension and the size of a dynamic network is given by $N \sim w^4$ (the power $4$ comes from the $4$ transverse dimensions of $d_c$). Since $w \sim
t^{1/2}$ we conclude that:
\begin{equation}
  t \sim N^{1/2}.
\label{t_N_relation_equation}
\end{equation}
Therefore for a dynamic network of size $N$ at
criticality, $P_s(t)$ decays exponentially after a time $t_\times$, with $t_\times
\sim N^{1/2}$ \cite{exponentialy_decays}.
Figure~\ref{DNFullAndSurvivelity}(b) presents simulation results for the
survivability of the giant cluster in a dynamic network at criticality.
The figure shows that, for different values of $N$ and $t > t_{\times}$,
$P_s(t) \sim t^{-1}$, as expected from DP in infinite dimensions. The
exponential decay for $t_{\times} > N^{1/2}$ can also be seen, in
agreement with Eq.~(\ref{t_N_relation_equation}). The inset of
Fig.~\ref{DNFullAndSurvivelity}(b) shows the collapse of survivability data after scaling by $N^{1/2}$, supporting
again Eq.~(\ref{t_N_relation_equation}).

The size of the giant component as a function of $N$ is derived by substituting
Eq.~(\ref{t_N_relation_equation}) in the DP relation $S(t) \sim t^2$.
\begin{equation}
S(N) \sim N.
\label{S_N_relation_equation}
\end{equation}
Figure~\ref{ClusterSizeFig}(b) presents simulation results illustrating this scaling relationship, as well as a corresponding relationship for static networks, where $S(N)$ is known to scale as $S(N) \sim N^{2/3}$ \cite{ER,Bollobas}.
$P_s(t)$ for static networks is also known to decay exponentially after a time
$t_{\times} \sim N^{1/3}$. The two systems clearly have different behavior
and properties at criticality, and thus belong to two
different universality classes.

\begin{figure}[h]
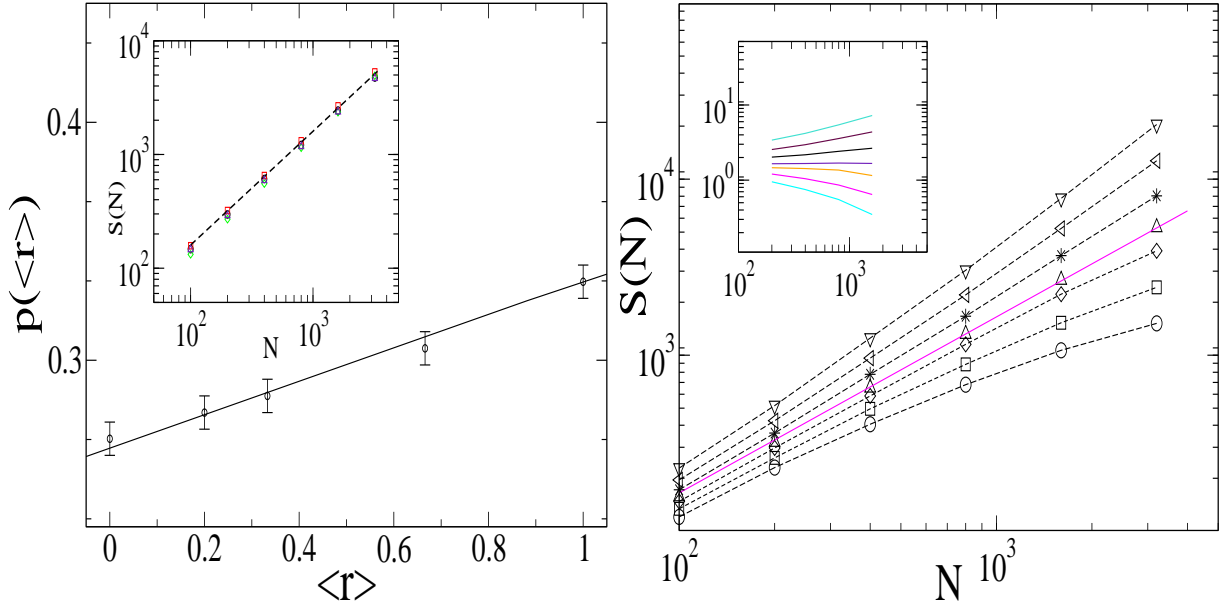

\begin{center}
\epsfig{file=p_c.eps,height=8cm,width=8cm}
\epsfig{file=fan.eps,height=8cm,width=8cm}
\end{center}
\caption{(a) Simulation fit to the formula for $p_c$ as a function of
  the rate $\langle r \rangle$ at which the links change.  The value of
  $p_c$ was calculated by finding the value of $p$ for which $S(N)$
  gives a straight line.  In the inset we show the value of $S(N)$ at
  $p_c$ for several different values of $\langle r \rangle$.  In all
  cases the slope equals $1$ as predicted by the critical exponent for
  dynamic networks.  (b) Simulation results for $S(N)$ in a network with
  $\langle k \rangle = 3$ and $\langle r \rangle = 2/3$ are presented
  for different values of $p$ showing that the network undergoes a phase
  transition at $p_c$.}
\label{Pc_formula}
\end{figure}

Next we show that the behavior of dynamic networks at criticality is
universal and independent of the rate in which the links are changed
[inset of Fig.~\ref{Pc_formula}(a)]. We do find, however, that the
critical concentration, $p_c$, for which the phase transition occurs
depends on the average lifetime of the links [Fig.~\ref{Pc_formula}(a)].
The formula for $p_c$ on Erd\H{o}s-R\'{e}nyi dynamic networks is given
by \cite{note3}:
\begin{equation}
p_c \equiv  \langle k \rangle^{-\langle r \rangle}(\langle k \rangle
+c(k))^{1-\langle r \rangle}
\label{critical_concentration}
\end{equation}
where $\langle r \rangle$ is the average rate with which the links change
and $\langle k \rangle$ is the average degree.  Fig.~\ref{Pc_formula}(b) presents simulation results for $S(N)$ at
different values of $p$ indicating the network undergoes a phase transition at
some critical value of $p$.  The inset of Fig.~\ref{Pc_formula}(a) shows
that for several different values of $r$ $S(N) \sim N$ at $p_c$, as
expected due to the universality of the critical exponent in dynamic
networks. The agreement of our simulations with the formula for $p_c $
[Eq.~(\ref{critical_concentration})] is shown in Fig.~\ref{Pc_formula}(a).

\begin{figure}[h]
\begin{center}
\epsfig{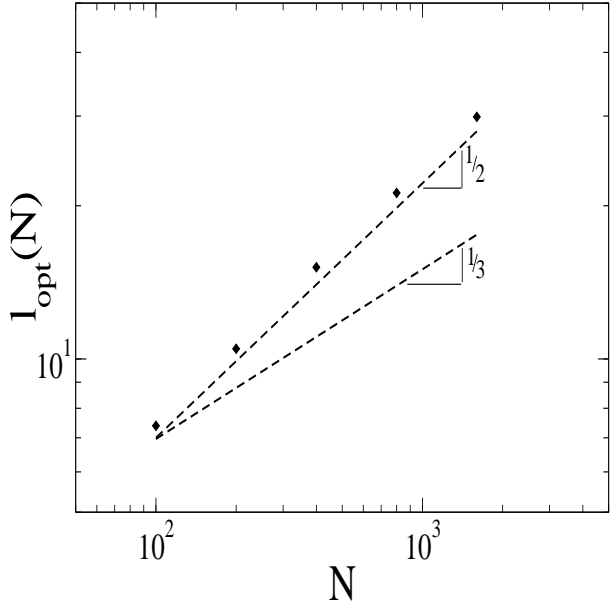}
\end{center}
\caption{The optimal path for strong disorder scales as $\ell(N) \sim
  N^{1/2}$ in dynamic networks compared to $\ell(N) \sim N^{1/3}$ in
  static networks.}\label{OptimalPathFig}
\end{figure}

The correspondence to DP can also predict the general scaling of the optimal
path in a dynamic network with a broad distribution of disorder. In a
network where weights are assigned to links, the optimal path between
any two nodes is defined as the path along which the sum of the weights
is minimal. In the limit of a broad distribution of disorder,
Ref.~\cite{Braunstein} has shown that, at criticality, the optimal path exists mainly
along the giant cluster.  Therefore for static networks
the optimal path length scales with the average distance between nodes
on the percolation cluster: $\ell_{\rm opt} \sim N^{1/3}$. In our
dynamic network model the average distance between nodes on the
percolation cluster scales as $\langle \ell \rangle \sim N^{1/2}$, suggesting that in dynamic ER
networks the optimal path scales as
\begin{equation}
\ell_{\rm opt} \sim N^{1/2}.
\label{OptimalPathEq}
\end{equation}
Figure~\ref{OptimalPathFig} shows simulation results for the optimal
path length in a dynamic network compared to a static network. The results for dynamic
networks are in full agreement with Eq.~(\ref{OptimalPathEq}).

What makes the results in a dynamic network so different from the static
case?  The difference lies in the number of available
configurations. While in static networks the percolation cluster is
composed from paths built from $N$ network nodes, in dynamic
networks the network is represented by $N^{3/2}$ nodes \cite{Note2}. The
evolution of the network over time generates many more possible
configurations, enabling the percolation cluster to become much larger.
Substituting $N' = N^{3/2}$ in the percolation cluster formula for
static networks, $S(N)\sim N^{2/3}$, yields $S(N')\sim S(N^{3/2}) \sim
N$, which further confirms our results for $S(N)$ in dynamic
networks.

The explanation for the optimal path being longer in directed
networks is that it can include loops. For example, an optimal path
reaching some node $A$ at $t = t_1$ may find it necessary to return to
that node at $t_2 > t_1$ if at $t_2$ node $A$ is more optimally
connected to the destination node. In a static network the optimal path
does not include loops since any link connecting node $A$ at $t_2$ was
also available at $t_1$ therefore any loop will only increase the total
weight.

Representing a dynamic network as a directed network
[Fig.~\ref{DNFullAndSurvivelity}(a)] composed of $N^{3/2}$ nodes allows
the ``same'' node to be counted more then once in the percolation
cluster, therefore requiring that the distinct number of nodes
on the percolation cluster also scale with $N$. To determine the number
of different nodes of the original network in a component of size $M$ on
the directed network, consider the following argument: The links between
consecutive layers of the directed network are chosen
randomly. Therefore, each link leads to a random node in the original
network independently and with uniform distribution. The probability to
reach a new node by following a link, assuming that $D$ nodes have
already been visited, is $1-D/N$. The expected number of distinct nodes $E(D)$
reached after $\lambda$ links have been followed from the starting node is therefore
$E(D_\lambda)/N=E(D_{\lambda-1})+E(1-D_{\lambda-1}/N)$. This reduces to
$E(D_\lambda)=1+(1-1/N)E(D_{\lambda-1})$ which indicates that for large
$M$
\begin{equation}
{E(D_M)\over N}=1-\left(1-\frac{1}{N}\right)^M\approx
1-e^{-M/N}\;.
\end{equation}
Thus, when the size of a component in the directed network is of order
$N$ a finite fraction of the visited nodes are new and the size of the
induced component on the original network is also of order $N$.

In summary, we introduced a model for dynamic networks which was solved
by a comparison with directed percolation in infinite dimensions. The DP
longitudinal axis is mapped to the time axis along which the dynamic
network evolves.  We showed that dynamic networks exhibit different
properties and critical exponents near criticality. Therefore they belong
to a different universality class than static networks.  While in static
networks $S(N)$, the size of the giant component at criticality, scales
as $S(N) \sim N^{2/3}$, in dynamic networks $S(N) \sim N$. Even though
the properties of dynamic networks are universal and independent of the
rate $r$ at which the links are changed, the critical concentration,
$p_c$, for which the phase transition occurs depends on $r$.  We also
showed that the optimal path in dynamic networks scales as $\ell_{\rm
  opt} \sim N^{1/2}$, compared to $\ell_{\rm opt} \sim N^{1/3}$ in
static networks.

\bigskip

\noindent
We thank the Israel Science Foundation,
Yeshaya Horowitz Association and The Center for Complexity Science,
EU Daphnet project, DOE and ONR for financial support,
and N. Madar and R. Bartsch for discussions.


\begin{thebibliography}{99}
\bibitem{exponentialy_decays} Reversing the relation from DP we get $t(w)
\sim w^2$, implying that if we limit the distance $w$ (instead of the
conventional way in DP of limiting the time $t$) we will obtain a cutoff
in the known critical exponents after time $t_\times \sim w^2$. The
equivalent for networks would be a cutoff after a crossover time
$t_\times \sim N^{1/2}$ (based on Eq. (\ref{t_N_relation_equation})).
\bibitem{note3}
For Erd\H{o}s-R\'{e}nyi static networks the percolation threshold is
given by $p_c \equiv \langle k \rangle^{-1}$. Since at criticality
the average number of new nodes reachable from a given node approaches 1,
$p\langle k \rangle=1$.  However, for percolation on our dynamic
network model where a node can be revisited since the links may change,
there are two limits.  First, for networks having links with a long
average lifetime, a node reached in a percolation process is connecetd
to an additional link (through which it was reached) on top of the
$\langle k \rangle$ links to which an average node is connected. Such a
node has a higher effective average degree that depends on the
probability $c(\langle k \rangle)$, to return to a previous node.  The
second limit is for networks changing constantly. In this case the
degree of a node reached in a percolation process is a random variable
with no memory, and its effective average degree remains $\langle
k\rangle$.  Therefore, the value of $p_c$ in dynamic networks depends on
the rate at which the links change and includes a combination of both
limits.  A good approximation for the probability of a node visited in a
percolation process to be revisited is obtained when considering the
first order contribution of more than one path to return to the original
node after two steps. This leads to $c(\langle k \rangle) \approx
(1-p^2)^{\langle k \rangle}$. The theoretical line in
Fig.~\ref{Pc_formula} was calculated according to this approximation.
\end{thebibliography}
\end{document}